\documentstyle[11pt]{article}
\input epsf

\begin{document}
\baselineskip 18pt

{\ }\hskip 4.7in SWAT 101 %put here preprint number
\vspace{.5in} 

\begin{center} {\bf QUANTISATION OF GLOBAL ISOSPIN IN THE SKYRME CRYSTAL} 
\baselineskip 13pt{\ }\\ 
{\ }\\ W.K. Baskerville\footnote{Work supported by PPARC.} \\ {\ }\\  
Physics Department\\
University of Wales, Swansea\\
Singleton Park\\
Swansea SA2 8PP, U.K.
\end{center} 

\begin{center}
March 1996\end{center}
\vspace{10pt}

\begin{quote}\baselineskip 13pt
  \noindent{\bf ABSTRACT:} A quantisation of chiral symmetry within
  the Skyrme crystal is carried out. The definition of global isospin
  in the crystal is explored, and found to be ambiguous. However, the
  state corresponding to a neutron crystal is identified, and the
  leading quantum correction to the classical mass is computed. The
  results are compared to those of Klebanov for a crystal of whole
  skyrmions. 
\end{quote}
\baselineskip 18pt

\section{Introduction}
\label{sec:intro}

The Skyrme model \cite{Skyrme1} has had considerable qualitative
success in describing both single nucleon properties and the
nucleon-nucleon interaction \cite{ANW,JackRho,Jacksons}. This has
prompted speculation as to whether the model might not also provide a
reasonable description of dense nuclear matter, such as may exist in
the interior of a neutron star. The idea of using a skyrmion crystal
for this purpose was first raised by Klebanov \cite{Kleb}. There is
some debate as to whether or not an ordered crystalline state is
energetically preferable to a disordered neutron superfluid at high
densities \cite {PandSmith}. Klebanov considered a simple cubic array
of skyrmions, appropriately rotated to ensure maximal attraction
between each skyrmion and its six nearest neighbours. Subsequent
investigations \cite{WBJ,Walhout} of this crystal revealed a phase
transition: at high densities, the crystal becomes a bcc array of
half-skyrmions (this was first realised by Manton and Goldhaber
\cite{GM}, who also identified an additional symmetry). The energy
minimum occurs in this high-density phase. Different crystal
symmetries were then tried \cite{JackVerb}. The lowest energy
configuration known consists of a simple cubic lattice of
half-skyrmions \cite{KS,Castetal}. While it cannot rigorously be
proved that this is indeed the lowest energy possible in the Skyrme
model, it seems very likely that it is. The energy obtained is only
3.8\% above the unreachable topological lower bound \cite{Skyrme1}. We
therefore refer to it as ``the'' Skyrme crystal.

Before the Skyrme crystal can be used to describe nuclear matter in a
neutron star, its global isospin must be quantised to ensure
electrical neutrality. It is generally agreed that a proper treatment
of the Skyrme model as a quantum field theory is extremely difficult.
Instead, a semi-classical quantisation is usually performed
\cite{ANW}, whereby the classical degrees of freedom of a given mode
are treated as collective coordinates. thus reducing the model to a
finite-dimensional quantum mechanics. The conventional wisdom is that
$6N$ degrees of freedom are required to describe a system containing
$N$ nucleons (the same number as would be required to describe 6
widely separated skyrmions). For the Skyrme crystal, global isospin
rotations should give the largest single quantum correction to the
classical mass. The global rotation of an infinite crystal requires
infinite energy; however a global isospin rotation requires only a
finite amount of energy {\em per baryon\/}. Klebanov calculated the
isorotational energy of his crystal, but this has so far been
neglected for the minimum energy Skyrme crystal. There are some
unresolved problems: skyrmion crystals predict too high a density for
nuclear matter, and the remainder of the kinetic energy may be
sufficient to unbind the crystal. It is therefore one of the aims of
this letter to see how Klebanov's results are modified for the true
Skyrme crystal. The classical Skyrme crystal is already known to have
a higher binding energy and a lower density than Klebanov's crystal.
We find that this tendency is even more heavily emphasised after the
quantisation of global isospin, so that it would seem very unlikely
that the Skyrme crystal could become unbound. Also, this (rather
limited) quantisation already leads to a $25\%$ correction to the
classical density. The quantisation of the remainder of the $6N$ modes
(which probably correspond to soft isospin and vibrational waves) is
an extremely difficult problem, but the results of the present
calculation indicate that tackling it may prove worthwhile.

We also desire to investigate the meaning of global isospin in an
infinite crystal. Isospin is conventionally defined in the Skyrme
model as an SO(3) rotation of the pion fields only. However, the full
symmetry of the Skyrme Lagrangian is somewhat larger: chiral ${\rm
  SU(2)}_{\rm L} \times {\rm SU(2)}_{\rm R} \cong {\rm SO(4)}$. This is
usually broken for finite energy configurations by the necessity of
setting the field to a constant value at spatial infinity
(conventionally $U \rightarrow 1$ as $r \rightarrow \infty$). However,
this condition does not apply to an infinite crystal. Also, it has
been argued \cite{Forketal,JackVerb} that the high density
half-skyrmion phase of the Skyrme crystal corresponds to a restoration
of chiral symmetry. In this case, there is no natural way to select
the diagonal subgroup corresponding to isospin. This would not matter
if all the fields transformed in the same way under the crystal point
groups. Unfortunately, however, they do not: one field is singled out.
The spectrum obtained by quantising global isospin is therefore
dependent on whether or not this field is included in the diagonal
subgroup. A unique energy spectrum can only be obtained by quantising
the full chiral (SO(4)) symmetry. However, the interpration of isospin
in these energy states is ambiguous, if indeed it can be meaningfully
defined at all. This problem was not considered by Klebanov for his
crystal (though it should also arise), as his calculation had a
particular definition of isospin built in from the start. The present
letter therefore represents the first attempt to address these issues.

\section{Calculations}
\label{sec:calc}

In dimensionless units, the Skyrme Lagrangian density is
\cite{skyrme2} 
\begin{equation}
  {\cal L} = \frac{1}{2}\, Tr(L_{\mu} L^{\mu}) + \frac{1}{16}\,
  Tr([L_{\mu}, L_{\nu}][L^{\mu}, L^{\nu}]),               \label{Ldef} 
\end{equation}
where $L_{\mu} = U^{\dagger} \partial_{\mu} U$ and $U = \sigma +
i\underline{\pi}.\underline{\tau}$ is the SU(2)-valued scalar field
($\tau_{i}$ are the Pauli matrices). In these units the topological
lower bound is $12 \pi^{2}$ per baryon.

We will begin by considering standard SO(3) isospin, and then
generalise to $SO(4)$ at the most convenient point. A global isospin
rotation of a Skyrme field $U$ is defined
\begin{equation}
 U(x) \mapsto A U(x) A^{\dagger},
  \label{su2isospin}
\end{equation}
where $A$ is an arbitrary $SU(2)$ matrix. This corresponds to an SO(3)
rotation of the pion fields $ \pi(x) \mapsto D(A) \pi(x)$, where
$D(A)$ is the $SO(3)$ matrix associated to $A$ via $D(A)_{ij} =
\frac{1}{2} \mbox{Tr}(\tau_{i}A\tau_{j}A^{\dagger})$. This
transformation is now allowed to depend on time
\begin{equation}
 U(\underline{x}, t) = A(t) U_{0}(\underline{x}) A^{\dagger}(t),
  \label{timedep}
\end{equation}
giving rise to kinetic terms in the Lagrangian. The components of $A$
are often treated directly as collective coordinates \cite{ANW}, but
we will instead follow the procedure of \cite{MLS} by defining the
body-fixed angular velocity $\omega$ for iso-rotations to be
\begin{equation}
 \underline{\omega}.\underline{t} = A^{\dagger}\dot{A},
  \label{omegadef}
\end{equation}
where $t_{i} = -\frac{1}{2} \tau_{i}$. The momentum conjugate to the
angular velocity $\omega$ will then be the body-fixed angular momentum
in iso-space. 

Substituting (\ref{timedep}) and (\ref{omegadef}) into the Lagrangian
density (\ref{Ldef}), the kinetic energy of the Skyrme crystal is
\begin{equation}
 T = \frac{1}{2}\, V_{ij}\, \omega_{i}\, \omega_{j}
  \label{Tso3}
\end{equation}
where
\begin{eqnarray}
 V_{ij}&  = & \int d^{3}x \left\{ \frac{1}{4} \mbox{Tr}\left( U^{\dagger}
 [\tau_{i}, U] U^{\dagger} [\tau_{j}, U] \right) \right. \nonumber \\
\ & \ & \ \ \ \ \ \ \ \ \ \ \ \ \ \ + \left. \frac{1}{16}
 \mbox{Tr}\left( [U^{\dagger}[\tau_{i}, U],\, U^{\dagger} \partial_{k}U]
 [U^{\dagger}[\tau_{i}, U],\, U^{\dagger} \partial_{k}U] \right)
\right\}  
  \label{Vdef}
\end{eqnarray}
is the isospin inertia tensor. 

Equation~(\ref{Tso3}) can now be generalised to SO(4) rotations of the
fields
\begin{equation}
  T = \frac{1}{2} V_{(ij)(kl)} \omega_{(ij)} \omega_{(kl)} ,
                                                         \label{Tso4}  
\end{equation}
where each of the indices runs from 0 to 3. Each of the pairs $(ij)$
is antisymmetric under the interchange of the two indices, while the
matrix $V$ is symmetric with respect to its `double' indices. We
choose to label $V$ by the pairs $(01)$, $(02)$, $(03)$, $(23)$,
$(31)$ and $(12)$, in that order. Most elements of $V_{(ij)(kl)}$ can
be calculated using the SO(3) formula~(\ref{Vdef}). The few remaining
elements are the cases where all four indices are different. They can
be computed by considering the case of motion of constant (but
different) velocity in two orthogonal planes, and selecting out the
cross-terms.

The form of the inertia tensor $V$ is strongly constrained by the
symmetry of the Skyrme crystal. There are two kinds of point about
which the Skyrme crystal has cubic symmetry: the centres of the
half-skyrmions (where the baryon density is peaked), and the points
where the `corners' of the deformed half-skyrmions meet (where the
baryon density is zero). The field transformations associated with
either point group are sufficient to define the crystal. Kugler and
Shtrikman \cite{KS} found the crystal fields by defining one point
group, writing down the most general Fourier expansion consistent with
this symmetry, and then determining the coefficients numerically by
minimising the energy. In an independent study, Castillejo et al.\ 
\cite{Castetal} discovered that right at the energy minimum, the
crystal fields are extremely well approximated by analytic formulae
\begin{eqnarray}
  \sigma & = & \sin \alpha \sin \beta \sin \gamma \nonumber \\
  \pi_{1} & = & \cos \alpha \sqrt{1 - \frac{1}{2} \cos^{2}
    \beta - \frac{1}{2} \cos^{2} \gamma + \frac{1}{3} \cos^{2} \beta
    \cos^{2} \gamma } 
                                                 \label{crystalfields} 
\end{eqnarray}
and cyclically for $\pi_{2}$ and $\pi_{3}$. $\alpha = \frac{\pi
  x}{L}$, $\beta = \frac{\pi y}{L}$ and $\gamma = \frac{\pi z}{L}$,
where $L$ is the lattice parameter. These formulae are a three
dimensional analogue of an exact two dimensional solution for the non
linear $\sigma$ model. Since these formulae encapsulate the symmetry
of the crystal, which is the most important feature for our purposes,
we adopt them for convenience. The form given in
Equation~(\ref{crystalfields}) assumes the origin to be at the centre
of the second point group mentioned above, but the symmetry with
respect to the first point group can easily be found by applying the
translation $x_{i} \mapsto x_{i} + L/2$ to these fields.

The full cubic point group consists of forty eight elements, which can
be divided into ten equivalence classes, each corresponding to a
particular physical symmetry of a cube.  Associated to each element
$g$ of the cubic group, there is a linear transformation of the fields
${\cal D}(g)$, where ${\cal D}(g)$ is a $4 \times 4$ matrix. The
matrices ${\cal D}(g)$ define a 4-dimensional representation of the
group. Surprisingly, the field transformations associated with the two
point groups of the Skyrme crystal correspond to different
representations of the cubic group. The cubic group has ten
irreducible representations (irreps), four of which are 1-dimensional,
two 2-dimensional and four 3-dimensional. The representation formed by
the crystal symmetry (for either point group) is 4-dimensional, and
must therefore be reducible. For the first point group (about centres
of half-skyrmions), the representation can be decomposed into the
trivial 1-dimensional irrep and a 3-dimensional irrep (the latter
corresponding to the transformations of the Cartesian axes under the
cubic group). For the second point group, the representation
decomposes into two 1-dimensional irreps (one of which is the trivial
representation) and a 2-dimensional irrep.

The crystal fields display a higher degree of symmetry with respect to
the first point group than to the second. The first point group
therefore imposes stronger constraints on the form of the inertia
tensor (including the conditions imposed by the second). The
decomposition into a 1-dimensional and a 3-dimensional irrep singles
out one direction in isospace. There are thus two principal moments of
inertia, one for iso-rotations purely within the 3-dimensional irrep ($B$),
and one for iso-rotations which mix the two ($A$).

This analysis is confirmed by numerical computation.  The
principal moments of inertia depend on the lattice parameter $L$
\begin{eqnarray}
 A & = & 6.6667 L^{3} + 69.4944 L ,\\ \nonumber
 B & = & 9.3333 L^{3} + 74.9876 L .                \label{moments}
\end{eqnarray}
All spatial integrations were performed over one unit cell. This is
necessary to ensure that the full symmetry of the crystal is
exhibited. The values for the principal moments of inertia given above
should therefore be interpreted as being the moments {\em per unit
  cell\/}. A unit cell consists of eight half-skyrmions in a cube of
side length $2L$, centred at the origin of one of the point groups.

The inertia tensor is diagonal if and only if a single field (rather
than a linear combination of fields) is chosen to correspond to the
trivial irrep. At this point, we note that the labelling of the
inertia tensor $V_{(ij)(kl)}$ implies a choice of the diagonal
subgroup (SO(3) isospin). It is clear that the diagonal elements of
$V_{(ij)(kl)}$ are $A$ (3 times) and $B$ (also 3 times). However, the
{\em order} will vary according to whether the field chosen to
correspond to the trivial irrep is $\sigma$, or one of the pion
fields. This will affect the form of the classical Hamiltonian when it
is expressed in terms of left and right SU(2) operators. Since the
Lagrangian is chirally invariant, the quantised energy spectrums
should be the same. However, the interpretation of isospin may vary.
We will consider two cases to illustrate this point, assigning
$\sigma$ and $\pi_{3}$ respectively to the trivial irrep. The
resulting inertia tensors are
\begin{equation}
  V_{1} = \left( \begin{array}{ccc|ccc}      
                   A & 0 & 0 & 0 & 0 & 0 \\
                   0 & A & 0 & 0 & 0 & 0 \\
                   0 & 0 & A & 0 & 0 & 0 \\ \hline
                   0 & 0 & 0 & B & 0 & 0 \\
                   0 & 0 & 0 & 0 & B & 0 \\
                   0 & 0 & 0 & 0 & 0 & B
                 \end{array} \right)  , \ \ \ \ \ \ \ \ 
  V_{2} = \left( \begin{array}{ccc|ccc}
                   B & 0 & 0 & 0 & 0 & 0 \\ 
                   0 & B & 0 & 0 & 0 & 0 \\
                   0 & 0 & A & 0 & 0 & 0 \\ \hline
                   0 & 0 & 0 & A & 0 & 0 \\
                   0 & 0 & 0 & 0 & A & 0 \\
                   0 & 0 & 0 & 0 & 0 & B
                 \end{array} \right) ,               \label{vdiff} 
\end{equation}
where $V_{1}$ obviously corresponds to the case where $\sigma$
transforms according to the trivial irrep.

For the first case ($V = V_{1}$), the Lagrangian can be written
\begin{equation}
 L = \frac{1}{2} A \Omega^{2} + \frac{1}{2} B \omega^{2} - V .
                                                     \label{L1}
\end{equation}
Defining conjugate momenta
\begin{equation}
 L_{i} = \frac{\partial L}{\partial \omega_{i}}, \ \ \ \ \ \ \ \ 
 M_{i} = \frac{\partial L}{\partial \Omega_{i}},   \label{conjmom} 
\end{equation}
the Hamiltonian can be written
\begin{equation}
 H = \frac{1}{2A} {\bf M}^{2} + \frac{1}{2B} {\bf L}^{2} + V .
  \label{Hso41lm}
\end{equation}

At this point, we need to convert to operator form to follow the usual
canonical quantisation procedure. However, $M$ and $L$ are not good
quantum numbers, as ${\bf M}$ and ${\bf L}$ do not separately obey the
correct angular momentum commutation relations. The symmetry algebra
is
\begin{equation}
 [L_{i}, \, L_{j}] = i \epsilon_{ijk} L_{k}, \ \ \ \ \ \ \ \ [M_{i},
 \, M_{j}] = i\epsilon_{ijk} L_{k}, \ \ \ \ \ \ \ \ [L_{i}, \, M_{j}]
 = i \epsilon_{ijk} M_{k}.
  \label{lmcomm}
\end{equation}
This is the algebra of vector ($L_{i}$) and axial ($M_{i}$) $SU(2)$
transformations (as would be expected, given the labelling of the
inertia tensor). Introducing
\begin{equation}
 J_{i} = \frac{1}{2} (L_{i} + M_{i}), \ \ \ \ \ K_{i} = \frac{1}{2}
 (L_{i} - M_{i}),
  \label{jkdef}
\end{equation}
we find that these operators {\em do} obey the correct angular
momentum algebra
\begin{equation}
 [J_{i}, \, J_{j}] = i \epsilon_{ijk} J_{k}, \ \ \ \ \ \ \ \ [K_{i},
 \, K_{j}] = i\epsilon_{ijk} K_{k}, \ \ \ \ \ \ \ \ [J_{i}, \, K_{j}]
 = 0.
  \label{jkcomm}
\end{equation}
This is the chiral algebra of $SU(2)_{L} \times SU(2)_{R}$. Rewriting
the Hamiltonian in terms of ${\bf J}$ and ${\bf K}$, we obtain
\begin{equation}
 H = \left(\frac{1}{2A} + \frac{1}{2B}\right) \left( {\bf J}^{2}
 + {\bf K}^{2} \right)  + 2 \left(\frac{1}{2B} -
 \frac{1}{2A}\right){\bf J}.{\bf K} + V . 
  \label{Hso41}
\end{equation}
From~(\ref{jkdef}), ${\bf J} + {\bf K} = {\bf L}$, so that
\begin{equation}
 2 {\bf J}.{\bf K} = {\bf L}^{2} - {\bf J}^{2} - {\bf K}^{2}.
  \label{jkop}
\end{equation}

Consider a crystal containing a large number $n$ of unit cells. We
assume strictly periodic boundary conditions and ignore edge effects
(ie.\ we put the crystal in a periodic box). Following the usual
canonical quantisation procedure, we convert to operator form
\begin{eqnarray}
 E^{tot} & = & n M_{cl} + \frac{\hbar^{2}}{nA} \left( J^{tot}(J^{tot} +
 1) + K^{tot}(K^{tot} + 1) \right) \nonumber \\
 \ & \ & \ \ \ \ \ \ \ +\  \hbar^{2}\left( \frac{1}{2nB} -
 \frac{1}{2nA} \right) L^{tot}(L^{tot} + 1).
  \label{so4spectrum1}
\end{eqnarray}
where $M_{cl}$ is the classical mass of a unit cell. 

We now consider the case where $\pi_{3}$ corresponds to the trivial
irrep, so that the inertia tensor is $V_{2}$ of
Equation~(\ref{vdiff}). The Lagrangian is then
\begin{equation}
 L = \frac{1}{2} B (\Omega_{1}^{2} + \Omega_{2}^{2}) + \frac{1}{2} A
 \Omega_{3}^{2} + \frac{1}{2} A (\omega_{1}^{2} + \omega_{2}^{2}) +
 \frac{1}{2} B \omega_{3}^{2} - V  .                        \label{L2} 
\end{equation}
Using the same definitions of ${\bf L}$, ${\bf M}$, ${\bf J}$ and
${\bf K}$, the classical Hamiltonian is
\begin{eqnarray}
  H & = & \frac{1}{2B} (M_{1}^{2} + M_{2}^{2}) + \frac{1}{2A}
  M_{3}^{2} + \frac{1}{2A} (L_{1}^{2} + L_{2}^{2}) + \frac{1}{2A}
  L_{3}^{2} + V                                     \label{Hso42} \\
  \ & = & \frac{1}{B} ({\bf J}^{2} + {\bf K}^{2}) + \left(
  \frac{1}{2A} - \frac{1}{2B} \right) {\bf L}^{2} +
  \left( \frac{1}{A} -\frac{1}{B} \right) (J_{3}^{2} + K_{3}^{2} -
    L_{3}^{2}) + V ,      \nonumber                              
\end{eqnarray}
which on conversion to operator form (again for $n$ unit cells) gives
\begin{eqnarray}
 E^{tot} & = & n M_{cl} + \frac{\hbar^{2}}{nB} \left( J^{tot}(J^{tot} +
 1) + K^{tot}(K^{tot} + 1) \right) \nonumber \\
 \ & \ & \ \ \ +\ \hbar^{2}\left( \frac{1}{2nA} -
 \frac{1}{2nB} \right) L^{tot}(L^{tot} + 1) \nonumber \\ 
 \ & \ & \ \ \ \ \ \ + \hbar^{2}\left( \frac{1}{nA} -
 \frac{1}{nB} \right) (J_{3}^{tot\ 2} + K_{3}^{tot\ 2} - L_{3}^{tot\ 2}) . 
                                                 \label{so4spectrum2} 
\end{eqnarray}
At first sight, this spectrum appears totally different to that given
by Equation~(\ref{so4spectrum1}), but in fact the energy eigenvalues (and
their degeneracies) are exactly the same. 

The different forms~(\ref{Hso41}) and~(\ref{Hso42}) of the classical
Hamiltonian reflect the differing choices of diagonal subgroup (SO(3)
isospin). We can reduce the chiral Hamiltonians to SO(3) isospin
Hamiltonians by setting ${\bf J} = {\bf K}$. Equation~(\ref{Hso41})
then reduces to 
\begin{equation}
 H = \frac{1}{2B} {\bf L}^{2} + V ,                     \label{Hso31}
\end{equation}
which gives the energy spectrum 
\begin{equation}
 E^{tot} =  n M_{cl} + \frac{\hbar^{2}}{2nB} L^{tot}(L^{tot} + 1) .
\label{so3spectrum1}
\end{equation}
Equation~(\ref{Hso42}) reduces to
\begin{equation}
 H = \frac{1}{2A} {\bf L}^{2} + \left( \frac{1}{2B} - \frac{1}{2A}
\right) L_{3}^{2} +  V ,                     \label{Hso32}
\end{equation}
with the corresponding energy spectrum
\begin{equation}
 E^{tot} =  n M_{cl} + \frac{\hbar^{2}}{2nA} L^{tot}(L^{tot} + 1) +
 \frac{\hbar^{2}}{2} \left( \frac{1}{nB} - \frac{1}{nA} \right)
 (L_{3}^{tot})^{2} .
\label{so3spectrum2}
\end{equation}
These spectrums {\em are} genuinely different. However, they give the
same energy for an infinite neutron crystal. For a neutron crystal,
$L^{tot} =2n$, $L_{3}^{tot} = -2n$, remembering that there are four
baryons per unit cell. Letting $n \longrightarrow \infty$, the energy
per baryon (from either spectrum) in a neutron crystal is
\begin{equation}
 \frac{E}{B} = \frac{1}{4} M_{cl} + \frac{\hbar^{2}}{2B} .
  \label{eperbneutron}
\end{equation}
Note that the cancellation of all terms involoving the moment $A$ in
the spectrum~(\ref{so3spectrum2}) only occurs in this
limit. Effectively, the isospin becomes large enough to be treated
classically, and the motion is reduced to rotation about one isospin
axis.

\section{Results and Conclusions}
\label{sec:results}

To summarise, the energy spectrum obtained from the quantisation of
chiral symmetry is unique, but the interpretation of isospin in these
energy states is ambiguous. If one only considered the
spectrum~(\ref{so4spectrum1}), it would seem that all SO(4) energy
eigenstates were of definite isospin ($L^{tot}$), degenerate with
respect to $L_{3}^{tot}$. However, states of the same energy can be
seen to have different isospin in spectrum~(\ref{so4spectrum2}).
Worse, not all the energy eigenstates in this second case are also
eigenstates of isospin. The last term of the Hamiltonian~(\ref{Hso42})
mixes states of different isospin but the same third component (for
given $J^{tot}$ and $K^{tot}$).

These results are not altogether surprising. Since the Lagrangian is
chirally invariant, it would be a shock to discover that the quantised
energy spectrum was not. However, the chiral symmetry is slightly
broken by the cubic symmetry of the Skyrme crystal, in that the
representation formed by the fields is reducible. This singles out one
field as `special'. All the trouble with the interpretation of isospin
stems from the presence of `spin-orbit' type terms in the chiral
Hamiltonians. These would disappear if all the principal moments of
inertia were equal, as they would be if all the fields transformed in
the same way under the cubic point group of the crystal. 

One possible way to resolve this ambiguity is to consider a finite
piece of crystal. It could be argued that for any physical application
the crystal would indeed be finite, even if very large. A choice of
vacuum at spatial infinity would then have to be made, breaking chiral
symmetry in the usual way. Furthermore, to maintain compatibility with
the cubic symmetry of the crystal, $\sigma$ must be chosen to
transform as the trivial irrep of whichever point group is centred at
the origin. Unfortunately, the same field does not correspond to the
trivial irrep for the two different point groups of the crystal, so
that the diagonal subgroup would still depend on the choice of origin.
The argument is further weakened by the idea that the high density
half-skyrmion phase of the Skyrme crystal may correspond to a
restoration of chiral symmetry \cite{JackVerb,Forketal}, which
suggests that the crystal interior might not be affected by boundary
conditions at its edge.

\begin{figure}[tb]
  \begin{center}
    \leavevmode
      {\hbox
      {\epsfxsize = 6.5cm \epsffile{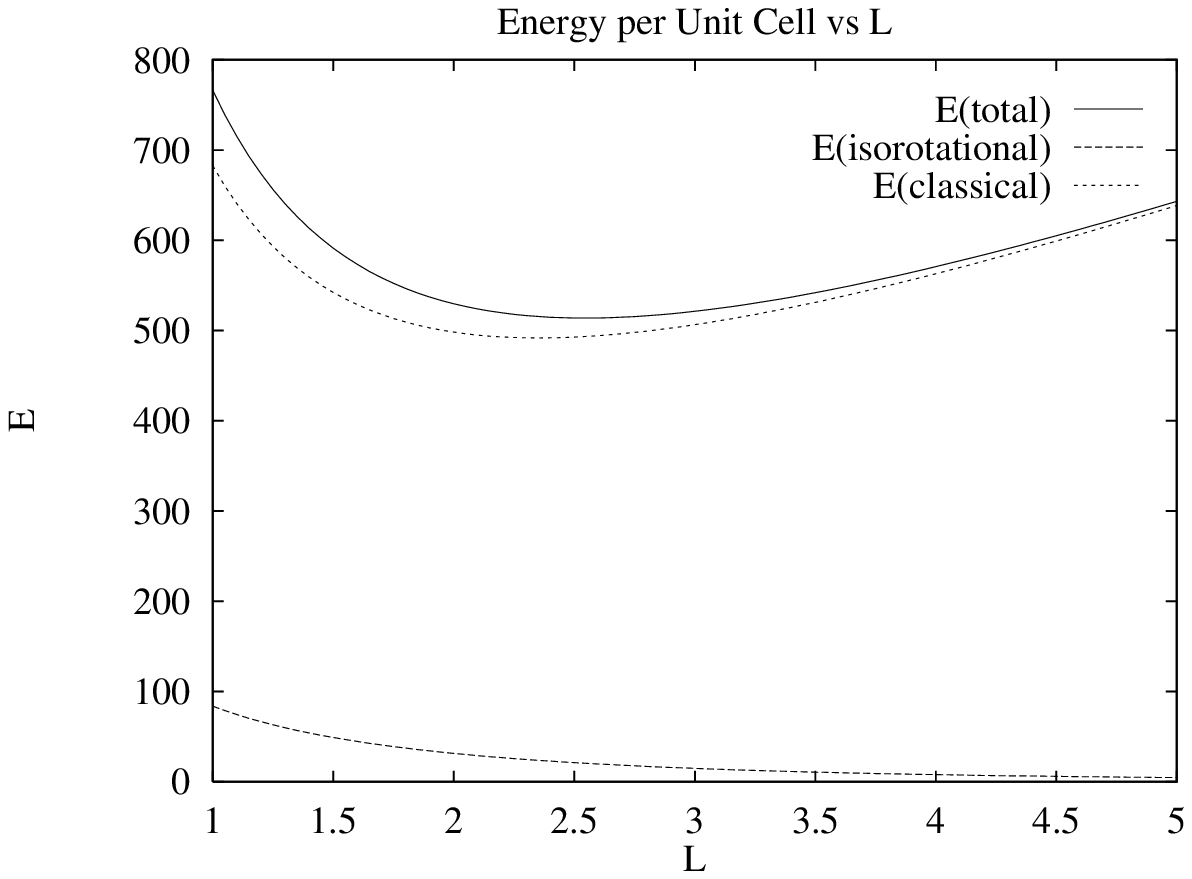}}
      {\epsfxsize = 6.5cm \epsffile{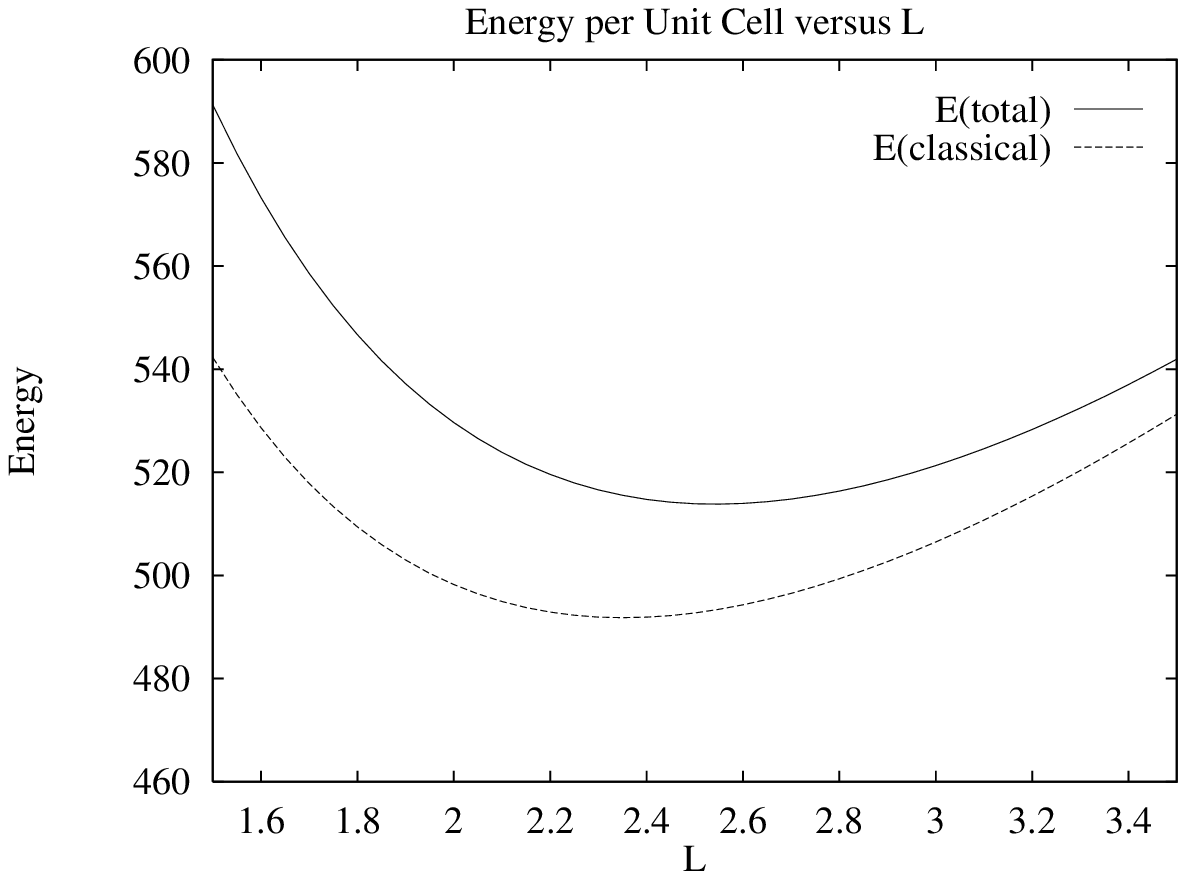}}
      }
  \end{center}
  \caption{Classical and isorotational energies per unit cell, and
    their sum, versus the lattice parameter L.}
  \label{fig:equant}
\end{figure}
Despite these difficulties, the energy of a neutron crystal would
appear to be uniquely defined by Equation~(\ref{eperbneutron}). We
will assume this to be true. The total, classical and isorotational
energies per unit cell are shown in Figure~\ref{fig:equant}. Part of
this graph is also displayed on a larger scale, to highlight the
difference in the value and position of the energy minimum, for the
classical and quantised Skyrme crystal. The quantisation of global
isospin has the effect of raising the minimum energy slightly from
491.81 to 513.84 (per unit cell) in our units: a difference of
approximately $4\%$. If the parameters of the Skyrme model are chosen
to fit the masses of the nucleon and the delta resonance, for zero
pion mass \cite{ANW}, then our energy units are equivalent to 5.92MeV,
our length unit equals 0.561 fm, and $\hbar = 59.4$ in our units. So,
translated into real units, the difference in energy is 32 $MeV$.
Perhaps more significant is the fact that the value of the lattice
parameter at the minimum increases from 2.35 to 2.54 in our units.
This means that the quantised crystal is almost $25\%$ less dense
(0.173 baryons/$\mbox{\rm fm}^{3}$ as compared to 0.218
baryons/$\mbox{\rm fm}^{3}$) than the classical Skyrme crystal.

\begin{figure}[tb]
  \begin{center}
    \leavevmode
    \epsfxsize = 13cm
    \epsfbox{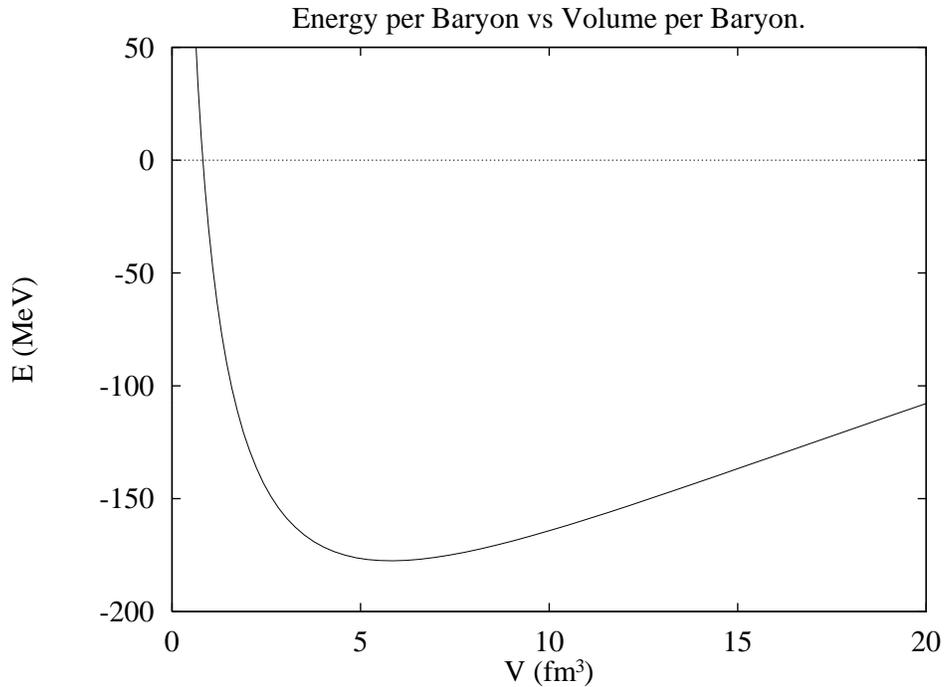}
  \end{center}
  \caption{Sum of classical and isorotational energies per
    baryon versus volume per baryon. The nucleon mass (938 MeV) has
    been subtracted.}
  \label{fig:kleb2}
\end{figure}

To facilitate comparison with the results of Klebanov \cite{Kleb} for
the crystal of whole skyrmions, we have also plotted the total energy
per baryon against the volume per baryon in standard units (see
Figure~\ref{fig:kleb2}). The true Skyrme crystal has a higher binding
energy, as would be expected. More surprisingly, it is also less dense
than the crystal considered by Klebanov.
As mentioned earlier, there is general agreement that the Skyrme
crystal predicts too high a density for nuclear matter. This
calculation seems promising in this respect, in that a crystal with
lower energy {\em and} density than that considered by Klebanov has
been found. The crystal is still too dense, but it must be remembered
that the quantisation carried out is only partial: only 6 out of the
$6N$ modes which should be included were actually considered (global
isospin, plus the zero momentum vibrational modes). This is
vanishingly few per baryon. It seems likely that quantisation of the
remaining modes will lead to further corrections to the density of the
crystal.  Allowing vibrational modes will almost certainly tend to
make the lattice parameter increase, and additional isospin modes
should also have the same trend, by analogy with global isospin.

It is difficult to know how much of the total kinetic energy per
baryon has been included in the present calculation.  Klebanov
suggests that the total kinetic energy per baryon should be of order
100 $MeV$ for his crystal, which could be enough to unbind it. It is
equally impossible to be sure that the true Skyrme crystal will be
bound, once the proper zero point energy is included. However, the
Skyrme crystal seems much more likely to be bound, since the binding
energy of the partially quantised neutron crystal is approximately
$20\%$ higher than for the Klebanov crystal. Overall, the trends
obtained from the limited quantisation performed here are extremely
promising. A full quantisation of the remaining modes, although
difficult, would appear to be interesting and worthwhile in the light
of these results.

\section*{Acknowledgements}
I would like to thank N. Manton for useful discussions of this work.


\begin{thebibliography}{99}
\bibitem{Skyrme1} 
T. H. R. Skyrme, {\em Proc. Roy. Soc.\/} {\bf 260} (1961) 127 
\bibitem{ANW} 
G.S. Adkins, C.R. Nappi and E. Witten, {\em Nucl. Phys. \/} {\bf B228}
(1983) 552  
\bibitem{JackRho} 
A.D. Jackson and M. Rho, {\em Phys. Rev. Lett. \/} {\bf 51} (1983) 751
\bibitem{Jacksons}
A. Jackson and A.D. Jackson, {\em Nucl. Phys. \/} {\bf A457} (1986)
687 
\bibitem{Kleb} 
I. Klebanov, {\em Nucl. Phys. \/} {\bf B262} (1985) 133
\bibitem{PandSmith}
V.J. Pandharipande and R.A. Smith, {\em Nucl. Phys. \/} {\bf A237}
(1975) 507  
\bibitem{WBJ}
E. W\"{u}st, G.E. Brown and A.D. Jackson, {\em Nucl. Phys. \/} {\bf
  A468} (1985) 450 
\bibitem{Walhout}
T.S. Walhout, {\em Nucl. Phys. \/} {\bf A484} (1988) 397
\bibitem{GM}
A.S. Goldhaber and N.S. Manton, {\em Phys. Lett. \/} {\bf B198} (1987)
231
\bibitem{JackVerb}
A.D. Jackson and J.J.M. Verbaarschot, {\em Nucl. Phys. \/} {\bf A484}
(1988) 419 
\bibitem{KS}
M. Kugler and S. Shtrikman, {\em Phys. Lett. \/} {\bf B 208} (1988)
491 
\bibitem{Castetal}
L. Castillejo, P.S.J. Jones, A.D. Jackson, J.J.M. Verbaarschot and
A. Jackson, {\em Nucl. Phys. \/} {\bf A501} (1989) 808
\bibitem{Forketal}
H. Forkel, A.D. Jackson, M. Rho, C. Weiss, A. Wirzba and H. Bang, 
\bibitem{skyrme2}
T.H.R. Skyrme, {\em Nucl. Phys. \/} {\bf 31} (1962) 556
\bibitem{MLS}
N.S. Manton, R.A. Leese and B.J. Schroers, {\em Nucl. Phys. \/} {\bf
  B422} (1995) 228

\end{thebibliography}
\end{document}